\documentclass[cameraready]{Interspeech}
\usepackage{url}
\usepackage{multirow}
\usepackage{subcaption}
\usepackage{xcolor}
\usepackage{booktabs}
\usepackage{multirow}
\usepackage{latexsym}
\usepackage{amsmath}  
\usepackage{amssymb}   
\usepackage{tikz} 
\usepackage{xcolor}
\usepackage[T1]{fontenc}
\usepackage[utf8]{inputenc}
\usepackage{microtype}
\usepackage{inconsolata}
\usepackage{graphicx}
\usepackage{ subcaption, algorithm2e, url, float, etoolbox, adjustbox, pgf,booktabs,verbatim}
\usepackage{xcolor,graphicx,colortbl}

\title{Bridging the Age Gap: Towards Detecting Neural Audio Codec Synthesized Elderly Speech Deepfake}

\author[affiliation={1}, equalcontribution]{Orchid}{Chetia Phukan}
\author[affiliation={2}, equalcontribution]{Girish}{}
\author[affiliation={3}, equalcontribution]{Mohd}{Mujtaba Akhtar}
\author[affiliation={1}]{Chi-Chun}{Lee}

\address{
    $^1$ BIIC Lab, NTHU, Taiwan, 
    $^2$ UPES, India, $^3$ VBSPU, India
}

\email{orchidchetiaphukan1@gmail.com, cclee@ee.nthu.edu.tw}


\usepackage{comment}

\begin{document}

\maketitle

\begin{abstract}

\noindent In this study, we introduce the Elderly CodecFake Detection (ECFD) task and release the Elderly-CodecFake (ECF) dataset in English and Chinese. We show that state-of-the-art CF detectors trained on previous benchmark CF datasets generalize poorly to elderly speech, revealing a critical vulnerability. We further hypothesize and demonstrate that multimodal foundation models (FMs) such as LanguageBind (LB) and ImageBind (IB) are more effective for ECFD due to their exposure to elderly content during cross-modal pretraining. Motivated by prior evidence that fusion of FMs enhances downstream performance, we explore fusion of FMs for ECFD. To this end, we propose \textbf{\texttt{BONSAI}}, a novel framework that employs Jensen–Shannon Divergence as the fusion mechanism. \textbf{\texttt{BONSAI}} with the fusion of LB and IB achieves an average EER (\%) of 1.66 and outperforms individual FMs as well as competitive SOTA baselines, establishing a new benchmark for the ECFD task.

\end{abstract}

\section{Introduction}

In recent years, the boundary between genuine and speech deepfakes has become increasingly blurred. Current text-to-speech (TTS) and voice-conversion (VC) models can generate speech utterances in nearly human-level realism. While these capabilities support valuable applications in assistive communication, human–computer interaction, and entertainment industry, they also enable serious misuse. They can be exploited for impersonation fraud, misinformation campaigns, and unauthorized replication of personal identities. Recognizing this threat, the research community has established dedicated benchmarks to drive the development of reliable countermeasures \cite{wu2015asvspoof, kinnunen2017asvspoof, wang2020asvspoof, liu2023asvspoof, wang2024asvspoof}. Early detection approaches relied on handcrafted features combined with traditional machine learning classifiers \cite{patel2015combining, patel2016cochlear, balamurali2019toward}. Succeeding them, researchers explored various deep learning approaches for detection of speech deepfakes leading to notable improvements \cite{tian2016spoofing, gomez2019light, jung2022aasist}. More recently, usage of large-scale speech foundation models have gained focus in the community; architectures such as Wav2vec2 and WavLM, trained on massive unlabeled corpora, have demonstrated strong transferability to speech deepfake detection \cite{martin2022vicomtech, tak2022automaticc, kawa23b_interspeech, guo2024audio, pimentel2024efficient, tran25b_interspeech, tran2025multi}. \par

\begin{figure}[hbt!]
    \centering
    \subfloat[]{%
        \includegraphics[width=0.23\textwidth]{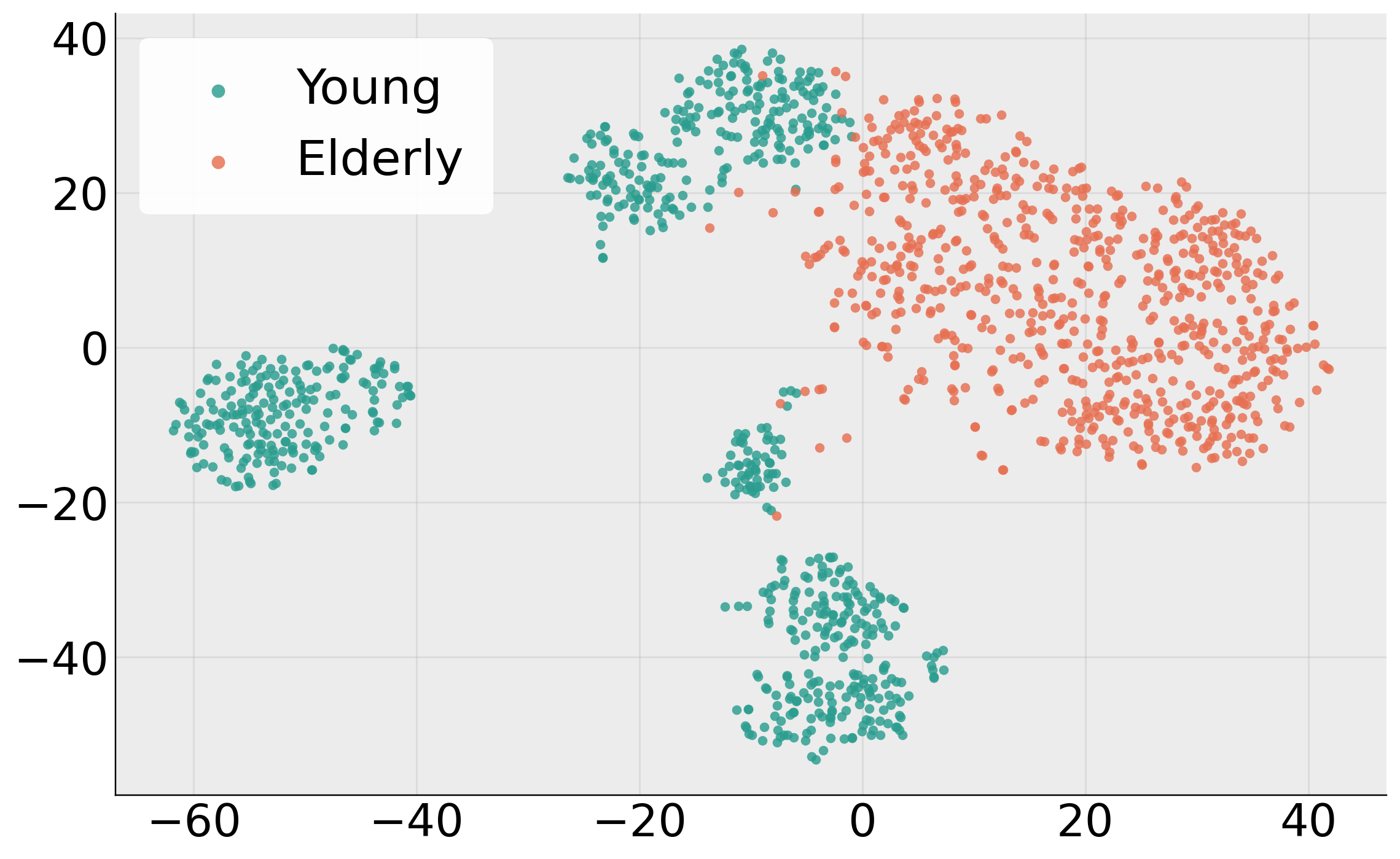}
    }
    \hfill
    \subfloat[]{%
        \includegraphics[width=0.23\textwidth]{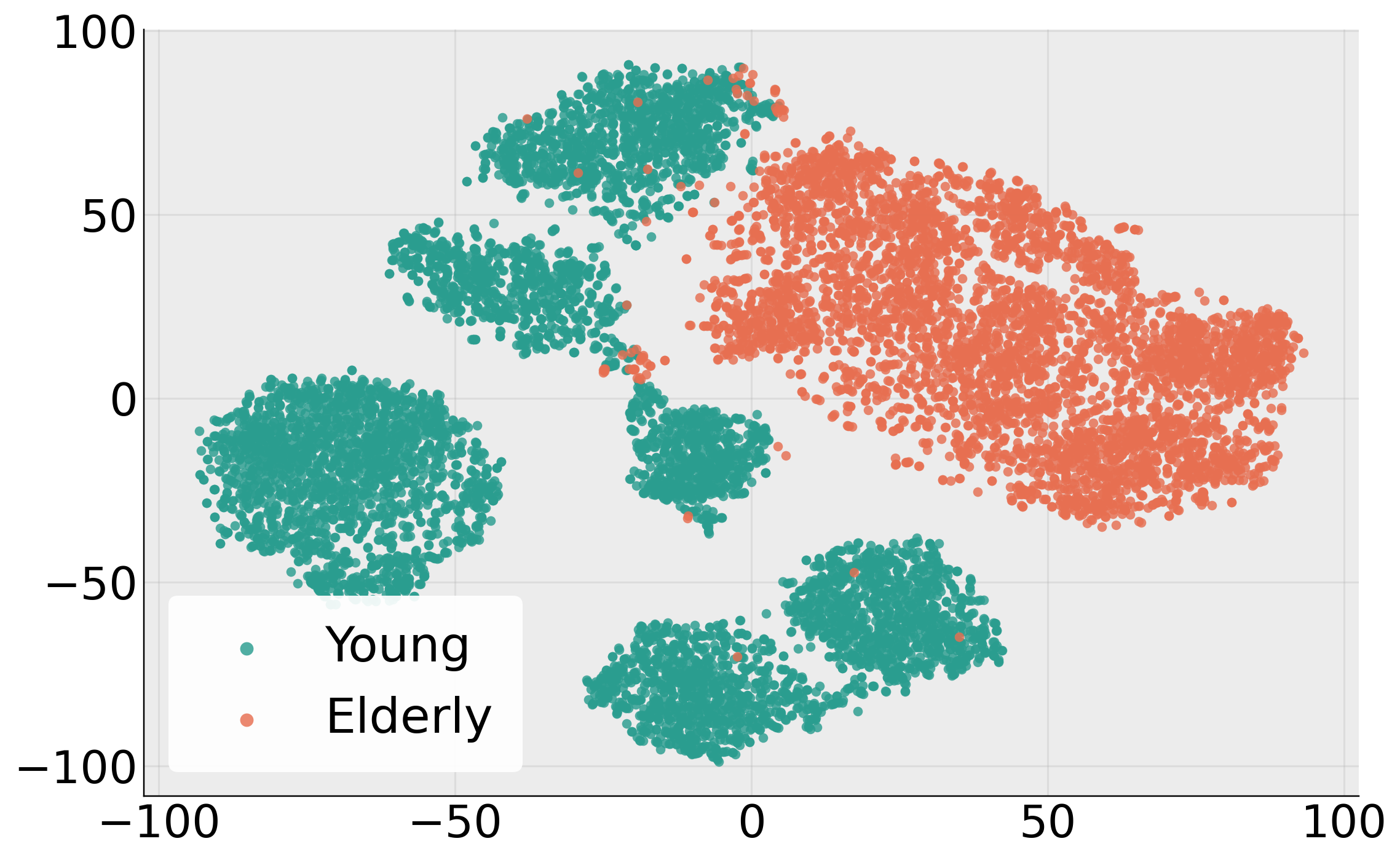}
    }
    \caption{t-SNE Plots for Real Speech (a) and CF Speech (b) samples for TIS Corpus\cite{maltezou2025human}; Clear separation of Young and Elderly Speech can be observed in both Real and CF scenarios}
    \label{fig:tsneelderlyyoung}
\end{figure}

However, prior studies have largely focused on detecting speech deepfakes generated via TTS, VC, or traditional vocoder-based models. With the rapid advancement of audio language models (ALMs), a new class of deepfakes has emerged, demanding dedicated countermeasures. These ALMs rely on neural audio codecs (NACs) for both encoding and synthesis (For example, AudioLM is built on Sounstream as NAC backbone \cite{borsos2023audiolm}), giving rise to the term CodecFakes (CFs). The first investigations into CF detection were conducted by Wu et al. \cite{wu24p_interspeech} and Lu et al. \cite{lu24f_interspeech}, who established initial benchmarks and detection strategies for this emerging threat. They showed that models trained on existing vocoder-based datasets fail when evaluated for detection of CFs. As such various researchers have come up with various approaches \cite{chen2025codecfake+, cui2025whiadd, xie2025codecfake}. Nevertheless, existing resources for CF detection are predominantly built on younger adult speakers, overlooking other demographics. This leaves older adults particularly vulnerable, as elderly speech exhibits distinct vocal traits—such as increased breathiness, reduced pitch stability, and irregular temporal patterns—that differ markedly from younger populations \cite{rojas2020does} (Figure \ref{fig:tsneelderlyyoung} (a)). Reflecting these differences, related fields such as speech emotion recognition have established dedicated benchmarks and challenges targeting elderly speech \cite{schuller20_interspeech, sogancoglu20_interspeech}. In contrast, research in CF detection has yet to address this demographic, leaving a critical gap in both datasets and methodologies. \par

To address this limitation, we introduce Elderly CodecFake Detection (ECFD) task (Figure \ref{fig:tsneelderlyyoung} shows the support for the need of ECFD task) and present the Elderly-CodecFake (ECF) dataset, comprising authentic and NAC-generated utterances from older speakers across diverse age bands, NAC backbones, and two languages—English and Chinese. We first evaluate state-of-the-art (SOTA) CF detectors trained on existing datasets \cite{lu24f_interspeech, wu24p_interspeech} and observe substantial performance degradation on NAC-generated elderly voices, revealing poor cross-demographic generalization. Motivated by the success of foundation models (FMs) for speech deepfake detection \cite{kawa23b_interspeech, tran2025multi}, we also explore FMs for ECFD. To our end, \textit{we hypothesize that multimodal FMs such as LanguageBind (LB) and ImageBind (IB)—are better suited for ECFD, as their cross-modal pretraining improve contextual understanding and implicitly capture age-related cues through exposure to visual contexts such as faces and scenes involving elderly individuals}. The effectiveness of leveraging multimodal FMs due to their cross-modality information prior for audio-centric tasks has also been demonstrated in related domains, such as non-verbal human vocalization emotion recognition \cite{phukan2025strong}. To validate our hypothesis, we conduct a systematic comparison between leading SOTA speech and multimodal FMs, and the results shows multimodal FMs as champion. Further, inspired by prior evidence that fusion of FMs improves performance for speech deepfake detection \cite{hetero2024}, we explore such fusion for ECFD and propose \textbf{\texttt{BONSAI}} (\textbf{B}ridging Fusi\textbf{ON} via Jen\textbf{S}en--Sh\textbf{A}nnon D\textbf{I}vergence), which employs Jensen-Shannon divergence as the fusion mechanism. \textbf{\texttt{BONSAI}}, applied to LB and IB, achieves the best performance, outperforming individual FMs and competitive SOTA baselines, establishing a new benchmark for ECFD. Our work lays the foundation for CF detection in underrepresented demographics.
 \par

\noindent \textbf{To summarize, the main contributions are as follows}: (i)  We formalize the novel task of ECFD and introduce the ECF dataset. (ii) We benchmark SOTA CF detectors trained on previous benchmark CF datasets and demonstrate a substantial performance degradation on elderly speech, revealing poor cross-demographic robustness. (iii) We show that multimodal FMs are best suited for ECFD by comparing SOTA speech and multimodal FMs. (iv) We propose a novel framework, \textbf{\texttt{BONSAI}}, which leverages Jensen-Shannon Divergence for fusion of FMs. \textbf{\texttt{BONSAI}} with LB and IB, achieves the best performance with 1.66 as average EER (\%), outperforming individual FMs and competitive SOTA baselines. We release the dataset and code here \footnote{\url{https://helixometry.github.io/ElderlyCodecFake/}}.

\section{Elderly Codecfake Dataset}
\label{ecf}
This section first describes the sources of real elderly speech data, followed by an overview of the NACs employed in this study. All data are obtained from publicly available corpora. Finally, we detail the pipeline used to generate ECF dataset.

\noindent \textbf{Real Elderly Speech Source}: \textbf{SeniorTalk} (E1) \cite{chen2025seniortalk}: It is a Mandarin conversational speech corpus specifically designed for super-aged seniors (Age 75 to 85), containing 55.53 hours of spontaneous dialogues from 202 speakers across 16 provinces. By prioritizing natural interaction rather than read speech, SeniorTalk offers a realistic benchmark for developing age-inclusive voice technologies. The dataset is splitted to training, validation, and test sets. \textbf{TIS Corpora} (E2) \cite{maltezou2025human}: The dataset comprises 1152 utterances from 96 speakers in English across diverse demographics, including younger (18–45) and older (60+) adults from White, Black, and South Asian backgrounds. It was designed to support inclusive speech technology by incorporating speakers from multiple age groups and racial identities.

\noindent \textbf{Neural Audio Codecs}: We follow previous works on CF detection works \cite{lu24f_interspeech, wu24p_interspeech} for selection of NACs and employ publicly released, reproducible NACs that reflect current SOTA generation pipelines. \textbf{Descript Audio Codec (DAC)} \cite{kumar2024high}: We utilize DAC models operating at sampling rates of 16\,kHz, 24\,kHz, and 44\,kHz. 
\textbf{EnCodec} \cite{defossez2022high}: We adopt the 24\,kHz and 48\,kHz variants in our experiments. 
\textbf{SoundStream} \cite{zeghidour2021soundstream}: The 16\,kHz configuration is employed. 
\textbf{Speech Tokenizer} \cite{zhang2024speechtokenizer}: We use the 16\,kHz model. 
\textbf{FunCodec} \cite{du2024funcodec}: The officially released 16\,kHz version is incorporated. 
\textbf{AudioDec} \cite{wu2023audiodec}: Models at 28\,kHz and 48\,kHz are used. 
\textbf{SNAC} \cite{siuzdak2024snac}: Multiple configurations at 24\,kHz, 32\,kHz, and 44\,kHz are employed. \textbf{MIMI} \cite{defossez2024moshi}: We use the 24\,kHz version. Considering all configurations, a total of fourteen NAC variants are used in this work.

\noindent \textbf{ECF Dataset Generation Process}: To build the ECF dataset, we adopt a structured pipeline inspired by following Wu et al. \cite{wu24p_interspeech} and Lu et al. \cite{lu24f_interspeech}, one of the foundational CF detection work. The procedure converts real elderly speech into NAC-generated counterparts using multiple NACs that constitute the core of modern ALM systems. We start from publicly available elderly speech datasets described above, where each original recording is treated as a real reference sample. For creating synthetic counterparts, every input speech sample undergoes a NAC encoding–decoding process. First, it is is transformed into a discrete latent sequence through the pre-trained encoder of a NAC, and this representation is subsequently reconstructed using the corresponding decoder. The resulting synthetic speech preserves linguistic content and speaker identity but contains NAC-induced distortions unique to each individual NAC, yielding realistic yet synthetic elderly speech. This procedure is applied across all fourteen NAC variants, producing parallel data such that each real speech utterance has a one-to-one synthetic counterpart for every codec type. For the SeniorTalk dataset, we follow the official data split. CF samples for the training and validation sets are generated using SNAC, DAC, EnCodec, SoundStream, SpeechTokenizer, and FunCodec (including their variants). For the test set, CF samples are generated using AudioDec, SNAC, and Mimi (including their variants). The TIS corpus contains speech samples from twelve elderly speakers. As no official split is provided, we perform a speaker-independent partition: eight speakers for training, two for validation, and two for testing. CF samples for all splits of TIS corpus are generated following the same procedure used for SeniorTalk. SeniorTalk and TIS contain 60029 and 720 real elderly utterances, respectively. In total, this results in 60749 real elderly speech utterances across both datasets. After generating CF data using fourteen NAC variants, the resulting dataset comprises 850486 elderly CF speech samples.


\section{Methodology}

This section presents the FMs employed in our study, followed by the downstream modeling approaches. We then detail the proposed framework, \texttt{\textbf{BONSAI}}.


\subsection{Foundation Models}
The FMs considered are SOTA in their respective benchmarks.   

\noindent \textbf{Mutlimodal FMs}: We select LanguageBind (LB) \cite{zhu2024languagebind} and ImageBind (IB) \cite{girdhar2023imagebind} as multimodal FMs. IB maps diverse modalities (image, audio, text, IMU, depth, thermal) into a shared image-centric embedding space via an InfoNCE objective, achieving strong cross-modal generalization without explicit paired supervision. LB similarly aligns modalities such as video, depth, audio, and infrared to a frozen language encoder using contrastive learning. 

\noindent \textbf{Speech FMs}: We employ Wav2vec2 \cite{baevski2020wav2vec}, WavLM \cite{chen2022wavlm}, and Whisper \cite{radford2023robust} as SOTA speech FMs. Wav2vec2 is included due to its demonstrated effectiveness in CF detection, particularly when combined with AASIST as a downstream model \cite{lu24f_interspeech}. We additionally incorporate WavLM, which has achieved strong performance across multiple speech processing tasks in the SUPERB benchmark and in speech deepfake detection \cite{guo2024audio}. Finally, we consider Whisper, a multilingual FM trained on 96 languages, unlike Wav2vec2 and WavLM which are primarily English-centric. Whisper has recently shown leading performance in speech deepfake detection \cite{hetero2024}. Wav2vec2 is a self-supervised model trained using a contrastive learning objective, while WavLM is also self-supervised but optimized through masked prediction and speech denoising tasks. In contrast, Whisper is pre-trained using a supervised multi-task learning framework. For all three FMs, we utilize their base variants. \par 

\noindent All audio samples are resampled to 16 kHz prior to feature extraction to ensure compatibility across FMs. We extract representations from the last hidden layer of each frozen FM by applying average pooling. For multimodal FMs, only the audio encoder branch is utilized. The resulting embedding dimensionalities are 768 for LB, Wav2vec2, WavLM; 1024 for IB; and 512 for the Whisper encoder.

\begin{figure}[hbt!] 
    \centering
    \includegraphics[width=0.4\textwidth, height=0.26\textheight]{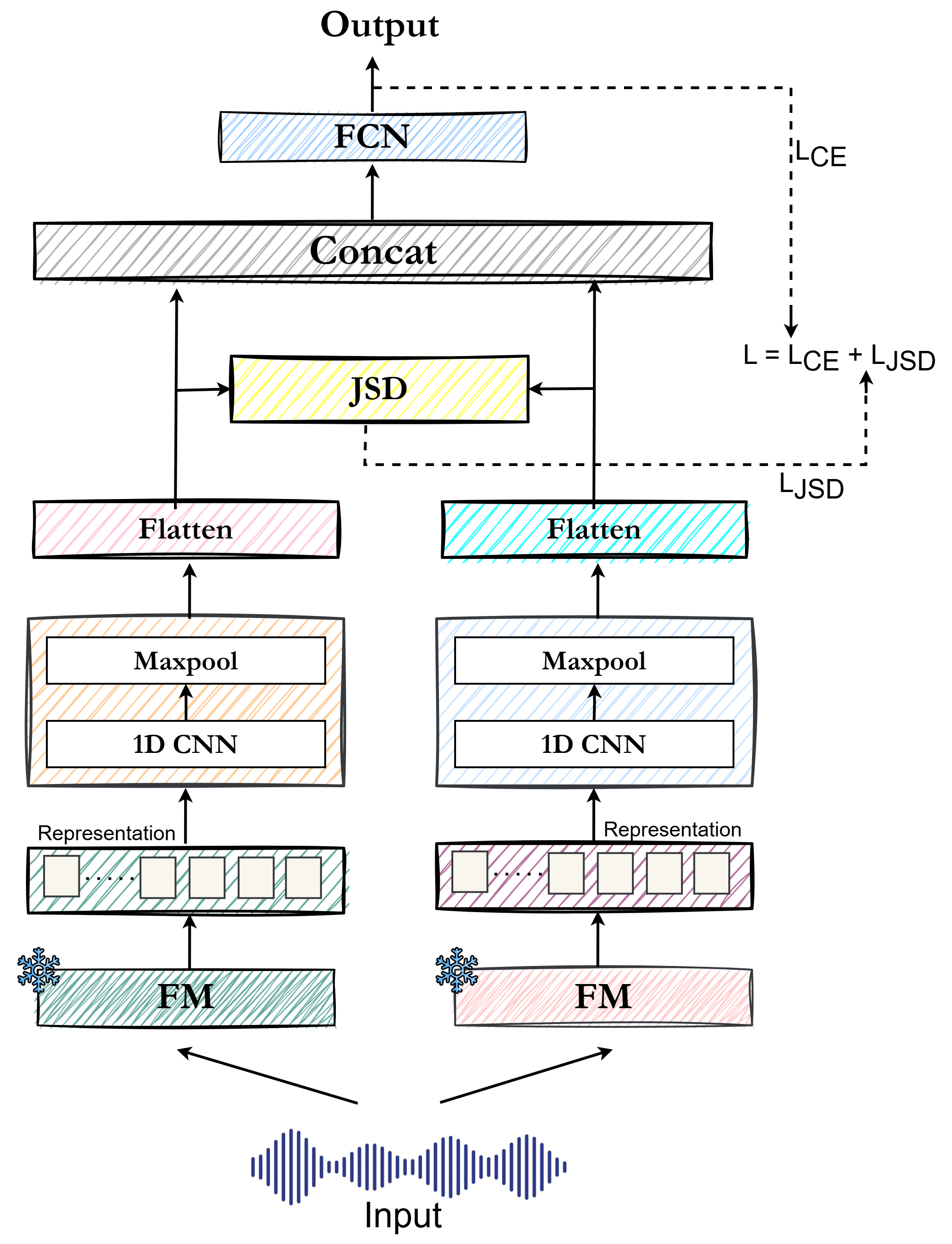} 
    \caption{Proposed Framework: \textbf{\texttt{BONSAI}}; JSD stands for Jensen-Shannon Divergence}
    \label{fig:archi_jsd}
\end{figure}

\subsection{Modeling}

Here, we detail the downstream modeling with individual FMs followed by discussion of the novel proposed approach for FMs fusion, \textbf{\texttt{BONSAI}}.

\noindent \textbf{Downstream Modeling with Individual FMs}: For downstream modeling, we adopt two widely used architectures for speech deepfake detection: AASIST \cite{jung2022aasist} and a convolutional neural network (CNN) \cite{hetero2024}. AASIST is a graph neural network–based architecture. In contrast, the CNN model consists of a1D-convolutional layer followed by max-pooling layer with a fully connected layer as output classifier. For both AASIST and CNN, we strictly follow the architectural configurations and implementation details described in Lu et al. \cite{lu24f_interspeech} and Phukan et al. \cite{hetero2024}, respectively, to ensure reproducibility and fair comparison.

\noindent \textbf{\texttt{BONSAI}}: The modeling architecture of \textbf{\texttt{BONSAI}} is illustrated in Figure \ref{fig:archi_jsd}. \textbf{\texttt{BONSAI}} leverages Jensen–Shannon Divergence (JSD) as a novel fusion mechanism for aligning FM representations. While JSD is primarily employed in multimodal learning for distribution matching and regularization \cite{xiao2018multi, sutter2020multimodal}, we repurpose it here as a novel loss function to explicitly align heterogeneous FM representations in a unified representation space. JSD is a symmetric and bounded measure of similarity between probability distributions which quantifies how individual distributions deviate from their shared mean distribution. We employ JSD because different FMs produce representations with distinct distributional characteristics due to differences in their pretraining objectives. Direct feature concatenation or linear fusion may therefore fail to account for distributional mismatch. By minimizing JSD between projected FMs representational space, \textbf{\texttt{BONSAI}} explicitly encourages distributional alignment in a stable and symmetric manner. The detailed modeling flow of \textbf{\texttt{BONSAI}} is given below: Representations from each FM are first passed through 1D-CNN layer with filter size 3 and we use 32 filters followed by maxpooling identical to those used with the CNN model mentioned in Downstream Modeling with Individual FMs. Then, the resulting features are flattened and projected to a shared dimensional space to ensure dimensional consistency and reduce computational overhead. Let ${e}_{a}$ and ${e}_{b}$ denote the projected flattened representations from two FMs. These vectors are normalized using softmax to obtain probability distributions $p$ and $q$. The JSD alignment loss is defined as: $
\mathcal{L}_{JSD} = \frac{1}{2} KL(p \parallel m) + \frac{1}{2} KL(q \parallel m) $ where $m = \frac{1}{2}(p+q)$ and $KL(\cdot)$ denotes the Kullback-Leibler (KL) divergence. Minimizing $\mathcal{L}_{JSD}$ encourages the two FM representations to capture complementary yet consistent information. Following this, the features are passed to a fully connected network (FCN) consisting of a dense layer with 120 neurons, followed by an output layer with a softmax activation function. The final training objective jointly optimizes classification ($\mathcal{L}_{CE}$) and alignment: $
\mathcal{L} = \lambda \mathcal{L}_{CE} + (1-\lambda)\mathcal{L}_{JSD} $ where $\lambda$ controls the balance between cross-entropy supervision and JSD-alignment. The number of trainable parameters in \textbf{\texttt{BONSAI}} ranges from 3.8M to 4.02M, depending on the dimensionality of the underlying FM representations.

\begin{table}[hbt!]
\centering
\scriptsize
\setlength{\tabcolsep}{5pt}
\renewcommand{\arraystretch}{1.15}
\begin{tabular}{l|c|cc}
\toprule
\multirow{2}{*}{\textbf{Model}} &
\multirow{2}{*}{\textbf{E1}} &
\multicolumn{2}{c}{\textbf{E2}} \\
\cmidrule(lr){3-4}
& & \textbf{Young} & \textbf{Elderly} \\
\midrule
AASIST           & 30.18 & 14.07 & 27.45 \\
Wav2vec2-AASIST  & \textbf{28.32} & \textbf{12.89} & \textbf{25.76} \\
\bottomrule
\end{tabular}
\caption{EER (\(\downarrow\)) in \%. Trained on Lu et al. \cite{lu24f_interspeech} and evaluated on ECFD test sets; E1: SeniorTalk, E2: TIS Corpus; The abbreviations are kept same for Table \ref{table2} and Table \ref{table3}}
\label{table1}
\end{table}


\section{Experiments}

\subsection{Training Details}

We train the models by combining the training sets from SeniorTalk and TIS, while validation and testing are performed separately on the respective validation and test splits of each individual dataset. All models are trained for 20 epochs using a learning rate of 1e-3 and a batch size of 32. We employ the Adam optimizer with cross-entropy loss for classification. For experiments involving \textbf{\texttt{BONSAI}}, the alignment weight $\lambda$ is set to 0.65, selected based on preliminary validation experiments. Dropout is employed during training to reduce overfitting, while class weighting is applied to handle class imbalance.

\subsection{Experimental Results}

We use Equal Error Rate (EER) as the evaluation metric, following prior work on speech deepfake detection and CF detection \cite{hetero2024, wu24p_interspeech, lu24f_interspeech}. \par
\noindent \textbf{Training on previous benchmark CF dataset \cite{lu24f_interspeech} and testing on ECFD (zero-shot)}: 
We train AASIST~\cite{wu24p_interspeech} and Wav2Vec2 with AASIST as the downstream classifier (Wav2Vec2-AASIST)~\cite{lu24f_interspeech} on the CF dataset introduced by Lu et al.~\cite{lu24f_interspeech}, which contains samples in both English and Chinese. These models represent SOTA CF detection approaches and are subsequently evaluated on the ECFD test set. The results are presented in Table~\ref{table1}. Furthermore, to analyze potential age-related performance bias, we generate CF samples for the younger speech subset of the E2: TIS corpus using the same neural audio codecs (NACs) employed to construct the ECFD test set. This enables us to assess whether models trained on the Lu et al.~\cite{lu24f_interspeech} CF dataset exhibit differential performance when evaluated on younger versus elderly speech samples. The results reveal clear cross-age performance degradation: the models achieve substantially lower EER on the younger subset, while the EER nearly doubles on the elderly subset. Notably, although the younger speech subset also constitutes an out-of-distribution condition relative to the training data, the models retain comparatively strong performance on younger speech than on elderly speech. This indicates that the observed degradation cannot be attributed solely to distribution mismatch, but is further attributed towards by age-related acoustic differences. Among the evaluated models, Wav2Vec2-AASIST demonstrates superior performance, likely due to its FM backbone, which provides better representations for CF detection. We also

\begin{table}[hbt!]
\centering
\scriptsize
\setlength{\tabcolsep}{5pt}
\renewcommand{\arraystretch}{1.15}
\begin{tabular}{l|ccc}
\toprule
\textbf{Model} & \textbf{E1} & \textbf{E2 (Elderly)} & \textbf{Avg} \\
\midrule
\multicolumn{4}{c}{\textbf{End to End: AASIST}} \\
\midrule
AASIST          & 14.54 & 13.66 & 14.10 \\
\midrule
\multicolumn{4}{c}{\textbf{Downstream: AASIST}} \\
\midrule
Wav2vec2 & 11.76 & 11.02 & 11.39 \\
WavLM  & 11.34 & 10.66 & 11.00 \\
Whisper  & 10.12 &  9.86 &  9.99 \\
IB       &  6.53 &  5.79 &  6.16 \\
LB       &  \cellcolor{yellow!25}\textbf{6.48} &  \cellcolor{green!25}\textbf{5.21} &  \cellcolor{yellow!25}\textbf{5.85} \\
\midrule
\multicolumn{4}{c}{\textbf{Downstream: CNN}} \\
\midrule
Wav2vec2 & 11.02 & 10.29 & 10.66 \\
WavLM    & 10.67 &  9.13 &  9.90 \\
Whisper  &  8.46 &  8.14 &  8.30 \\
IB       &  \cellcolor{green!25}\textbf{5.41} &  \cellcolor{yellow!25}\textbf{5.26} &  \cellcolor{green!25}\textbf{5.34} \\
LB       &  \cellcolor{blue!25}\textbf{4.81} &  \cellcolor{blue!25}\textbf{4.30} &  \cellcolor{blue!25}\textbf{4.56} \\
\bottomrule
\end{tabular}
\caption{Evaluation scores (EER (\(\downarrow\)) in \%) for training and evaluation on ECFD; Avg represents the average of EER across E1 and E2 (Elderly)}
\label{table2}
\end{table}

\noindent \textbf{In-domain Training and Evaluation on ECFD}: 
Table~\ref{table2} reports the in-domain results on the ECFD dataset. We first evaluate the AASIST baseline~\cite{wu24p_interspeech}, as it demonstrated better performance than prior baselines in Table~\ref{table1}. We then evaluate different FMs using both AASIST and CNN as downstream classifiers. The results show that multimodal FMs consistently outperform speech-only FMs, thereby supporting our hypothesis. To further analyze this behavior, we perform a qualitative analysis using t-SNE visualizations of the raw FM representations from LB and Wav2Vec2. The t-SNE plots reveal clearer separation and better clustering between real and fake classes for LB compared to Wav2Vec2. Furthermore, the CNN achieves stronger performance than AASIST while being more lightweight, whereas AASIST appears more prone to overfitting in this setting. In Table~\ref{table3}, we investigate whether combining FMs can further improve ECFD performance. We use simple concatenation as a baseline, employing the same architecture and training protocol as \textbf{\texttt{BONSAI}}, except without the JSD-based alignment loss. Overall, \textbf{\texttt{BONSAI}} consistently outperforms concatenation across all FM pairs, demonstrating that explicit representation alignment through JSD enables more effective exploitation of complementary information. The performance gains are modest for speech-only FM pairs but become more pronounced when fusing multimodal FMs or combining speech and multimodal FMs. Notably, combinations involving LB and IB achieve the strongest performance, indicating that multimodal FMs provide highly complementary representations for ECFD. We also evaluated KL divergence as an alternative alignment objective, given that JSD is derived from KL divergence. However, KL divergence showed less stable convergence and achieved inferior performance than \textbf{\texttt{BONSAI}}, with results comparable to simple concatenation. Consequently, we present concatenation as the baseline and omit KL divergence results due to space constraints.

\begin{figure}[hbt!]
    \centering
    \subfloat[]{%
        \includegraphics[width=0.23\textwidth]{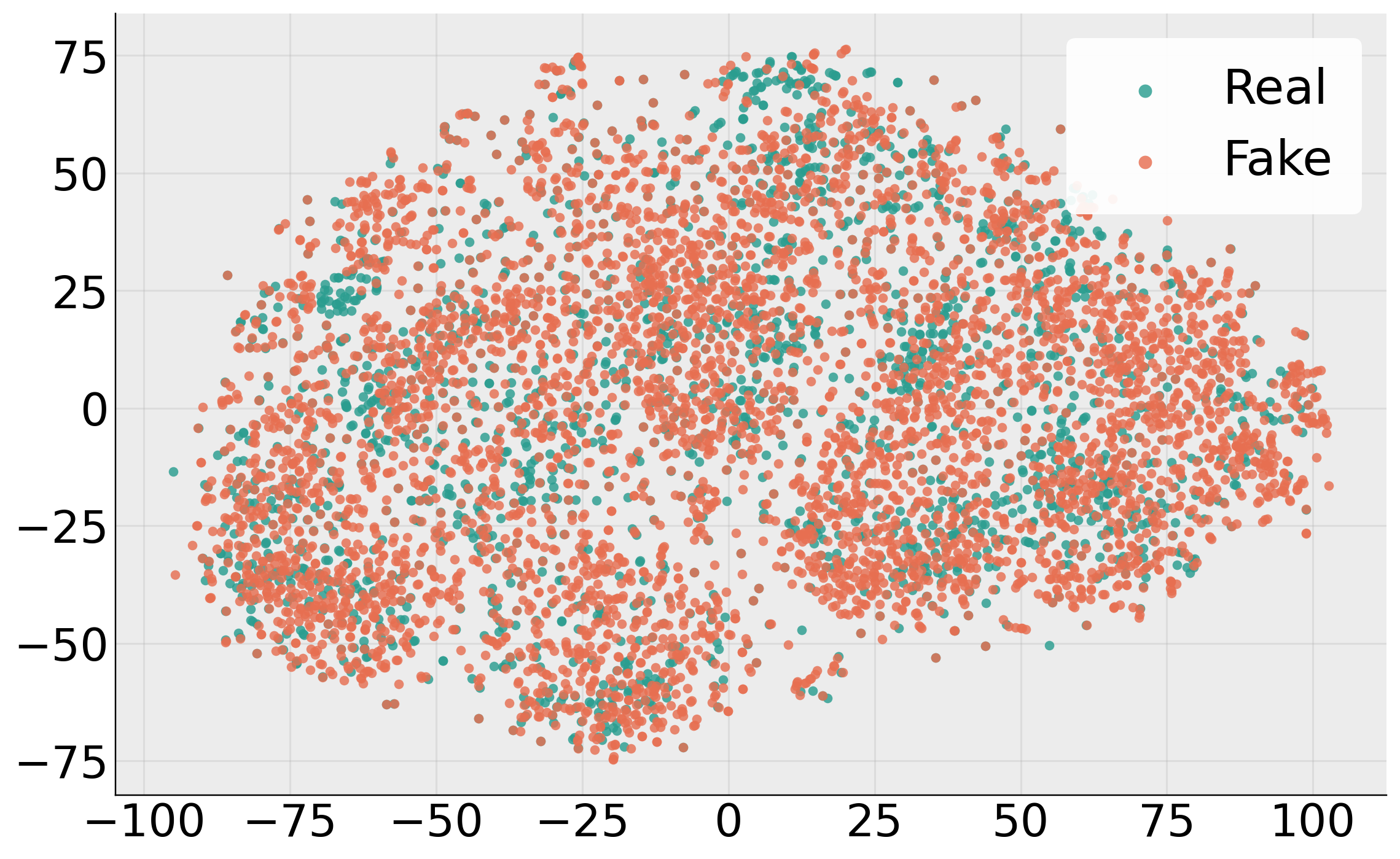}
    }
    \hfill
    \subfloat[]{%
        \includegraphics[width=0.23\textwidth]{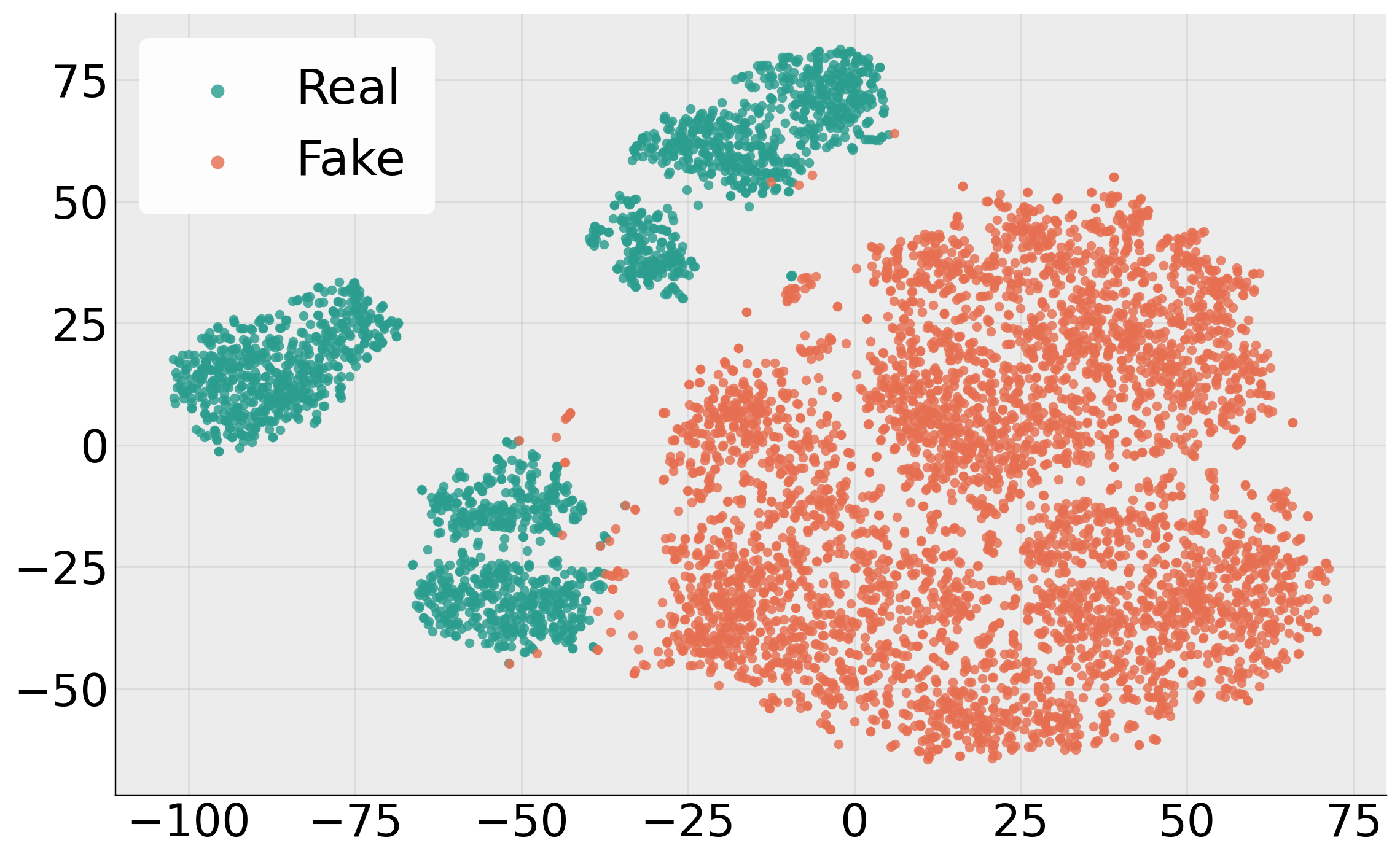}
    }


    \caption{t-SNE Plots (a) Wav2vec2 (b) LanguageBind}
    \label{fig:tsne}
\end{figure}

\begin{table}[!hbt]
\centering
\scriptsize
\setlength{\tabcolsep}{5pt}
\renewcommand{\arraystretch}{1.15}
\begin{tabular}{l|ccc|ccc}
\toprule
\multirow{2}{*}{\textbf{Pairs}} &
\multicolumn{3}{c|}{\textbf{Concatenation}} &
\multicolumn{3}{c}{\textbf{\texttt{BONSAI}}} \\
\cmidrule(lr){2-4}\cmidrule(lr){5-7}
& \textbf{E1} & \textbf{E2 (Elderly)} & \textbf{Avg}
& \textbf{E1} & \textbf{E2 (Elderly)} & \textbf{Avg} \\
\midrule
W2v2 + WL & 9.36 & 9.01 & 9.18 & 7.54 & 6.97 & 7.26 \\
W2v2 + Wh & 7.17 & 6.63 & 6.90 & 6.24 & 5.64 & 5.94 \\
W2v2 + IB & 5.20 & 4.79 & 5.00 & 4.87 & 3.22 & 4.05 \\
W2v2 + LB & \cellcolor{yellow!25}\textbf{4.47} & 4.22 & 4.35 & 4.00 & 3.76 & 3.88 \\
\midrule
WL + Wh   & 7.94 & 7.35 & 7.65 & 5.68 & 5.17 & 5.43 \\
WL + IB   & 5.12 & 4.72 & 4.92 & 4.36 & 4.09 & 4.23 \\
WL + LB   & 4.86 & 4.43 & 4.65 & 3.92 & 3.56 & 3.74 \\
\midrule
Wh + IB   & 4.51 & \cellcolor{yellow!25}\textbf{4.02} & \cellcolor{yellow!25}\textbf{4.27}
          & \cellcolor{yellow!25}\textbf{3.23} & \cellcolor{yellow!25}\textbf{2.95} & \cellcolor{yellow!25}\textbf{3.09} \\
Wh + LB   & \cellcolor{green!25}\textbf{3.78} & \cellcolor{green!25}\textbf{3.14} & \cellcolor{green!25}\textbf{3.46}
          & \cellcolor{green!25}\textbf{2.78} & \cellcolor{green!25}\textbf{2.36} & \cellcolor{green!25}\textbf{2.57} \\
\midrule
IB + LB   & \cellcolor{blue!25}\textbf{3.01} & \cellcolor{blue!25}\textbf{2.50} & \cellcolor{blue!25}\textbf{2.76}
          & \cellcolor{blue!25}\textbf{1.80} & \cellcolor{blue!25}\textbf{1.51} & \cellcolor{blue!25}\textbf{1.66} \\
\bottomrule
\end{tabular}
\caption{Evaluation scores (EER (\(\downarrow\)) in \%) for training and evaluation on ECFD; Avg represents the average of EER across E1 and E2 (Elderly); W2v2: Wav2vec2, WL: WavLM; Wh: Whisper}
\label{table3}
\end{table}

\noindent \textbf{Comparison to SOTA}: AASIST and Wav2vec2-AASIST represent SOTA architectures for CF detection~\cite{lu24f_interspeech, wu24p_interspeech}. From Table~\ref{table1} and Table~\ref{table3}, we observe that \textbf{\texttt{BONSAI}} with the fusion of LB and IB achieves the lowest average EER of 1.66\% (1.80\% on E1 and 1.51\% on E2 (Elderly)), thereby establishing a new SOTA for the ECFD task.

\section{Conclusion}

In this work, we introduced the ECFD task and the ECF dataset comprising English and Chinese speech. Our analysis demonstrated that existing SOTA CF detection models trained on prior benchmark datasets generalize poorly to elderly speech, revealing a critical robustness gap. Furthermore, we showed that multimodal FMs, such as LB and IB, provide more effective representations for ECFD. Building on this insight and motivated by the complementary nature of FMs, we proposed \textbf{\texttt{BONSAI}}, a novel fusion framework that leverages Jensen–Shannon Divergence for representation alignment. Experimental results demonstrated that \textbf{\texttt{BONSAI}} with LB and IB achieves an average EER of 1.66\%, outperforming individual FMs and competitive baselines, and establishing a new benchmark for the ECFD task. In future work, we plan to extend the ECF dataset to additional languages to further improve the robustness and generalizability of ECFD systems.

\section{Acknowledgement}

This work was supported by National Science and Technology Council (NSTC), Taiwan (grant\#: 115-2634-F-002-012).

{
\section{Generative AI Use Disclosure}
AI Assistants were utilized exclusively to enhance grammatical accuracy, clarity, and the overall readability of the manuscript. These tools did not contribute to the development of scientific concepts, data analysis, generation of results, or interpretation of findings. The authors assume full responsibility for the accuracy and integrity of the work.}

\bibliographystyle{IEEEtran}
\bibliography{mybib}

\end{document}